\documentclass[twocolumn]{revtex4-2}
\usepackage{graphicx}
\usepackage{dcolumn}
\usepackage{bm}
\usepackage{amsmath}
\usepackage{amsfonts}
\usepackage{amssymb}
\usepackage{color}
\usepackage{epstopdf}
\usepackage{siunitx}
\usepackage{float}
\usepackage{xcolor}
\providecommand{\U}[1]{\protect\rule{.1in}{.1in}}

\begin{document}
\title{Irreversibility of sudden processes: Compression versus expansion in perfect gases}
\author{Andr\'es Vallejo}
\author{Mat\'ias Osorio}
\affiliation{\begin{small} Instituto de F\'{\i}sica, Facultad de Ingenier\'{\i}a, Universidad 
de la Rep\'ublica, Montevideo, Uruguay\end{small}}
\date{\today}
\begin{abstract}
\noindent We focus on the study of the processes undergone by a perfect gas 
when the external pressure is suddenly modified. The analysis shows that, 
from the second law perspective, there is a non-evident asymmetry 
between the processes of compression and expansion between the same pressures, 
which is manifested 
through different levels of entropy production. If the system remains in 
contact with a thermal reservoir during the processes this asymmetry is 
remarkable, since while entropy grows linearly with the relative change 
of pressure during the compression, the relation between those variables 
is logarithmic during the expansion. 

\end{abstract}

\maketitle

\section{\label{sec:level1}Introduction}

Irreversibility is an universal feature of macroscopic physical processes. 
However, despite its ubiquity, understanding the causes of irreversible behaviour has 
represented a huge challenge, given the reversible character of the underlying 
microscopic laws, both in the classical and in the quantum regime. Although an 
universally accepted explanation of irreversibility is still lacking, this problem 
has played an important role in the development of physics \cite{Halliwell}. 

A system is said to undergo an irreversible process when it is impossible to 
return the system as well as the environment to their initial states. The heat 
exchange between 
bodies at different temperatures, abrupt expansions and the presence of friction 
are typically mentioned in thermodynamics textbooks as paradigmatic examples of 
irreversible processes \cite{vanWylen,Cengel,Zemansky,Callen}, but all processes 
that occur in practice are irreversible to a greater or lesser degree. It is clear 
that in the mentioned cases the system can be returned to its initial configuration, 
but at the expense of external resources that modify the state of the environment. 

Despite the fundamental role they play within the theory, the understanding of concepts such 
as reversibility, irreversibility or quasi-static process presents difficulties that, in our 
opinion, classic textbooks in the undergraduate level does not always address satisfactorily. 
In order to contribute to this issue, in this paper we focus on the study of thermodynamic 
irreversibility performing the comparison, from the second-law perspective, between 
the compression and expansion processes undergone by a perfect gas after a sudden 
change of the external pressure exerted on the system. 

Several works have addressed 
these topics using original approaches, in order to develop teaching strategies 
to be applied in undergraduate courses \cite{Anacleto2009,Miranda2008,Leff2018}. 
In particular, the reversible limit of the compression and expansion processes between 
two fixed pressures have been studied in Ref. \cite{Anacleto} for the case of an ideal gas 
in a thermally isolated piston-cylinder device. 
It was shown that if the processes are performed by placing the system in contact 
with $N$ work reservoirs at increasing (decreasing) pressures, the entropy variation 
of the universe tends to zero as the number of reservoirs $N$ goes to infinity. Here we 
analyse the opposite limit in which the processes are performed in one step, and our main 
interest is to compare these processes in terms of their departure from the reversible 
limit studied in \cite{Anacleto}.

Although many readers may argue that sentences like \textit{process A 
is more/less irreversible than process B} is nonsense and that processes either are 
reversible or are not, in practice, quantifying the level of irreversibility of a 
process is of paramount importance. It provides a measure of the work that could have 
been done and was lost due to dissipation sources, such as friction, irreversible 
chemical reactions, etc. In engineering thermodynamics, several measures of 
irreversibility have been defined and proved to be useful, such as \textit{exergy destroyed}, 
\textit{second-law efficiency}, or \textit{irreversibility} itself \cite{vanWylen,Cengel}. 
Since all of them are in one way or another linked to entropy production, we will use this 
quantity as a measure of the lost work due to the presence of irreversibilities in the process.

The outline of this paper is as follows. In Section II we compare the processes of 
irreversible compression and expansion for a perfect gas in contact with a thermal reservoir. 
The adiabatic case is analyzed in Section III, and some remarks 
and conclusions are presented in Section IV.

\section{System with diathermal walls}
Let us consider 1 mol of a perfect gas (an ideal gas with constant heat capacities 
$C_{P}$ and $C_{V}$), contained in a frictionless cylinder-piston device with 
diathermal walls (system A). Initially, the system is in thermal equilibrium with 
the environment (system B) at temperature $T_{1}$, and the pressure (due to the joint 
action of the atmosphere and the piston weight) is $P_{1}$. Let us consider the 
irreversible process suffered by the gas after abruptly increasing the weight over 
the piston in a constant amount such that it would be equilibrated by an internal pressure 
$P_2$. Although at first the gas temperature rises, it is clear that, due to the heat 
exchange through the walls, the gas will reach a new equilibrium state at pressure $P_2$ 
and at the original temperature $T_1$. In what follows, we evaluate the closeness of the process
to the reversible limit through the entropy production, which coincides with the 
global entropy variation.
 
The entropy change of the gas can be found integrating the Gibbs relation \cite{vanWylen}
	\begin{equation}\label{Gibbs}
	TdS=dH-VdP,
	\end{equation}
along a reversible trajectory linking the initial and final states. The simplest path in 
order to perform the integration is the isothermal one, for which $dH=0$ since the enthalpy 
of an ideal gas is a function of the temperature only \cite{vanWylen,Cengel}. Combining the 
Gibbs relation with the equation of state of an ideal gas, we obtain:  
	\begin{equation}\label{DS_gas_comp1}
	\Delta S^{A}_{1\rightarrow 2}=-\int_{P_1}^{P_2}\dfrac{R}{P}dP,
	\end{equation}
where $R$ is the universal gas constant. For reasons 
that will become clear shortly, we decide not to perform the integration in an explicit way.	
	
From the first law \cite{vanWylen}:
	\begin{equation}
	\Delta U^{A}_{1\rightarrow 2}=Q^{A}-W^{A},
	\end{equation} 
and noting that the internal energy of the gas is also a function of the temperature only, we can 
conclude that the heat exchanged equals the work done by the gas. Here we note that 
even if the pressure of the gas is not defined along the whole process, the external pressure 
is known and constant and, consequently, we can find the heat exchanged as \cite{Gislason}:
	\begin{equation}
	Q^{A}=\int_{V_{1}}^{V_{2}}P_{\text{ext}}dV=P_2\left(V_{2}-V_{1}\right).
	\end{equation}
The initial and final volumes can be expressed in terms of the pressures employing the 
equation of state for an ideal gas ($n$=1 mol):
	\begin{equation}\label{Eq.State}
		\begin{cases}
		V_{1}=RT_1/P_{1}\\
		V_{2}=RT_1/P_2,
		\end{cases}
	\end{equation}
and, from the above equations, we obtain that
	\begin{equation}
	Q^A=RT_{1}\left(1-\dfrac{P_2}{P_1}\right).
	\end{equation}
Considering the environment as an internally reversible heat reservoir at 
constant temperature $T_1$, we have that its entropy variation is
	\begin{equation}\label{DS_env_comp1}
	\Delta S^{B}_{1\rightarrow 2}=\dfrac{Q^B}{T_{1}}=\dfrac{-Q^A}{T_{1}}=\dfrac{R}{P_1}\left(P_2-P_1\right),
	\end{equation}
and finally, from Eqs. (\ref{DS_gas_comp1}), and (\ref{DS_env_comp1}), the entropy change 
of the universe for the sudden compression can be expressed as:
\begin{equation}\label{DS_univ_comp1_2}
\Delta S^{\text{Univ}}_{1\rightarrow 2}=\Delta S^{A}_{1\rightarrow 2}+\Delta S^{B}_{1\rightarrow 2}=\dfrac{R}{P_1}\left(P_2-P_1\right)- \int_{P_1}^{P_2}\dfrac{R}{P}dP.
\end{equation}

Now, let us consider the irreversible expansion undergone after removing the weight from 
the piston. Since the gas returns to the initial state, its entropy variation is opposite 
to that corresponding to the process $1\rightarrow 2$:
	\begin{equation}\label{DS_gas_exp1}
	\Delta S^{A}_{2\rightarrow 1}=\int_{P_1}^{P_2}\dfrac{R}{P}dP.
	\end{equation}

Once again the heat exchanged coincides with the work done, but in this case the expansion 
occurs under a different external pressure (the initial pressure):
	\begin{equation}\label{Q_exp}
	Q^{A}=W^{A}=\int_{V_{2}}^{V_{1}}P_{\text{ext}}dV=P_{1}\left(V_{1}-V_{2}\right).
	\end{equation}
Then, from Eqs. (\ref{Eq.State}) and (\ref{Q_exp}), we have that:
	\begin{equation}
	Q^{A}=RT_1\left(1-\dfrac{P_1}{P_2}\right),
	\end{equation}
and, consequently:
	\begin{equation}\label{DS_env_exp1}
	\Delta S^{B}_{2\rightarrow1}=\dfrac{-Q^{A}}{T_{1}}=-\dfrac{R}{P_2}\left(P_2-P_1\right).
	\end{equation}
Finally, from Eqs. (\ref{DS_gas_exp1}) and (\ref{DS_env_exp1}), the global entropy variation 
is:
	\begin{equation}\label{DS_univ_exp2_1}
	\Delta S^{\text{Univ}}_{2\rightarrow 1}=\int_{P_1}^{P_2}\dfrac{R}{P}dP-\dfrac{R}{P_2}\left(P_2-P_1\right).
	\end{equation}

\begin{figure}[!h]
	{\includegraphics[scale=0.7,angle=0, clip]{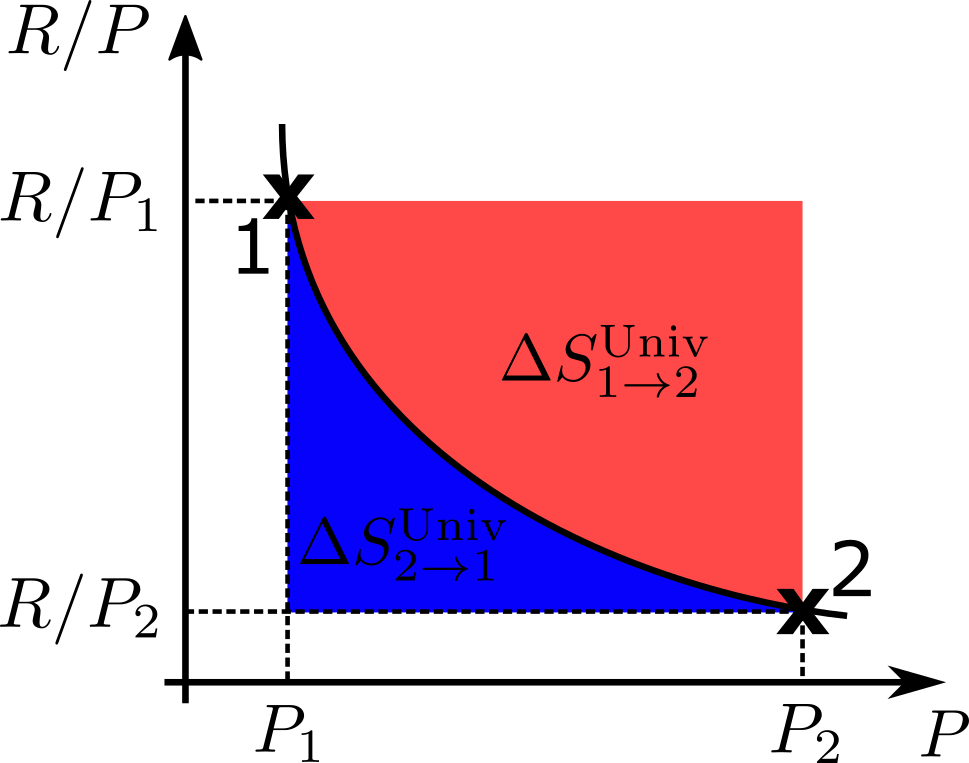}}
	\caption{Plot of $f(P)=R/P$ as a function of P. The entropy productions associated to 
	the compression and expansion processes between the pressures $P_1$ and $P_2$ can be 
	interpreted as areas in that diagram.}
	\label{fig1}
	\end{figure}

\begin{figure}[!h]
		{\includegraphics[scale=0.38,angle=0, clip]{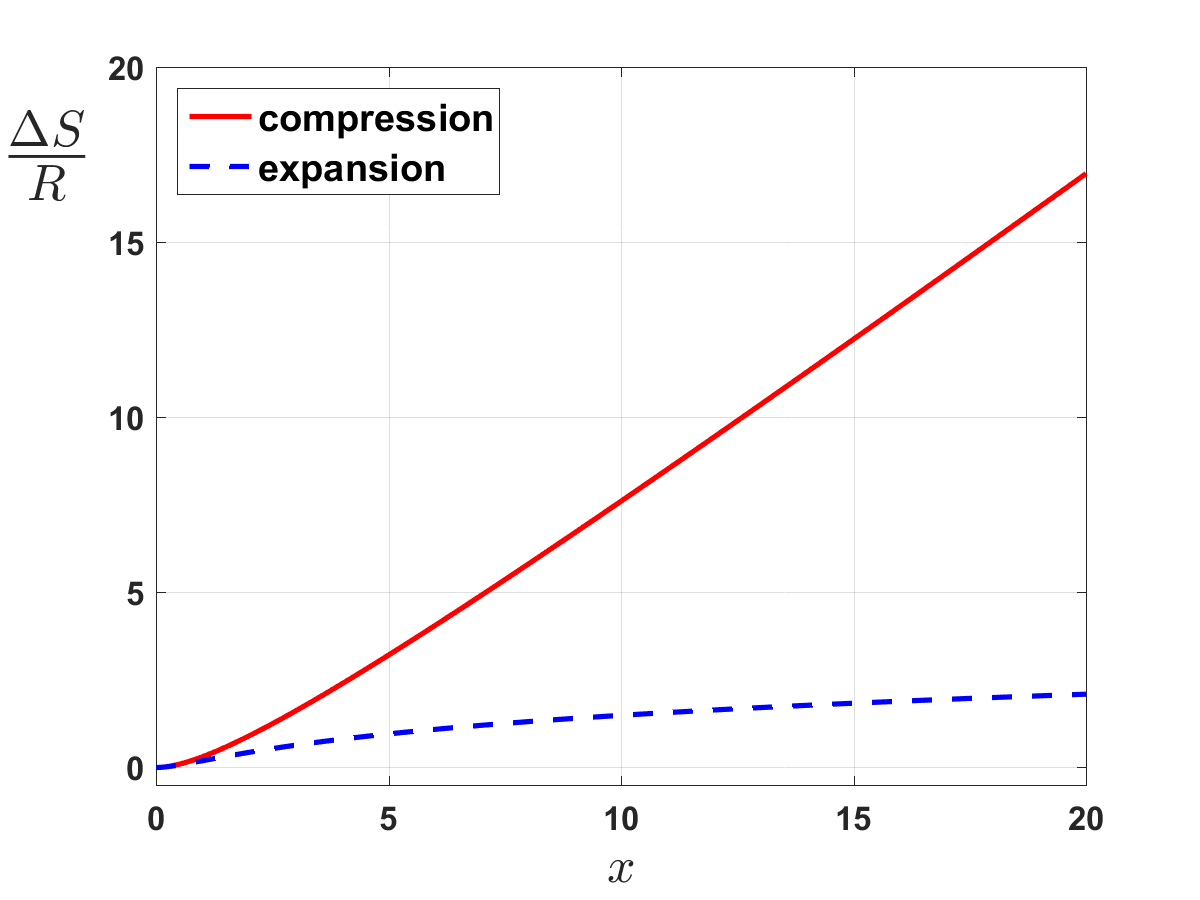}}
		\caption{Entropy production (in units of $R$) as a function of $x$ for the compression 
		(red, solid line) and the expansion (blue, dashed line) sudden processes. The gas exchanges 
		heat with its surroundings at temperature $T_1$.}
	 	\label{fig2}
	\end{figure}

It is relatively easy to perform an analytic study of the expressions (\ref{DS_univ_comp1_2}) 
and (\ref{DS_univ_exp2_1}) in order to compare both entropy productions. However, it is more 
illustrative to consider the diagram of Fig. (\ref{fig1}), in which we plot $R/P$ as a function 
of $P$. 
Regarding the 
compression process $1\rightarrow 2$, notice that the first term in Eq.(\ref{DS_univ_comp1_2}), 
$(R/P_1)\left(P_2-P_1\right)$, is the area of the rectangle whose base is the interval of the $P$ 
axis between $P_1$ and $P_2$, and whose height is $R/P_1$. Meanwhile, the term $\int_{P_1}^{P_2}R/P~dP$, 
is the area under the graph of the function $f(P)=R/P$ in the same interval. Consequently, 
the entropy production $\Delta S_{1\rightarrow 2}^{\text{Univ}}$ is represented by the upper
region in Fig. (\ref{fig1}). A similar reasoning allows us to quickly identify the entropy 
production in the expansion process as the lower region of the figure. Notice that 
the union of both regions is a rectangle with sides are parallel to the axes. Since this 
rectangle is crossed by the curve $f(P)$ by opposite vertices, the convex character of 
$f(P)$ implies that the compression always generate more entropy than the expansion 
between the same pressures. This asymmetry in irreversibility increases with the difference 
in pressures, as the construction of Fig. (\ref{fig1}) for a larger value of $P_2$ makes 
evident. In fact, from Eqs. (\ref{DS_univ_comp1_2}) and (\ref{DS_univ_exp2_1}), we can see 
that defining the relative change of pressure:
	\begin{equation}\label{x}
	x=(P_2-P_1)/P_1 = \Delta P/P_1,
	\end{equation}
Eq. (\ref{DS_univ_comp1_2}) adopts the form:
	\begin{equation}\label{DS_univ_comp1_2_2}
	\Delta S^{\text{Univ}}_{1\rightarrow 2}=R\left[x-\ln(1+x)\right],
	\end{equation}
presenting a linear behaviour with $x$ in the limit of large $x$, while, on the other hand
	\begin{equation}\label{DS_univ_exp1_2_2}
	\Delta S^{\text{Univ}}_{2\rightarrow 1}=R\left[\ln(1+x)-\dfrac{x}{1+x}\right],
	\end{equation}
so the entropy is produced logarithmically in the expansion process (see Fig. (\ref{fig2})).
	
In the opposite limit $(x\ll 1)$, the Taylor expansions of Eqs. (\ref{DS_univ_comp1_2_2}) and (\ref{DS_univ_exp1_2_2}) allow to show that
	\begin{equation}
	\Delta S^{\text{Univ}}_{1\rightarrow 2}\simeq\Delta S^{\text{Univ}}_{2\rightarrow 1}= \dfrac{R}{2}\left(\dfrac{\Delta P}{P_1}\right)^2+\mathcal{O}\left(\dfrac{\Delta P}{P_1}\right)^3,
	\end{equation}
so in both cases the entropy is produced quadratically with the relative change of pressure, 
a result that has been reported in Ref. \cite{Gupta}.

\section{System with adiabatic walls}

We proceed to compare the processes of compression and expansion in the limit in 
which the gas is thermally isolated (i.e. the cylinder is built with adiabatic walls). 
After adding the weight over the piston, the 
system again evolves irreversibly towards an equilibrium state at pressure $P_2$. 
But, in this case, the temperature of the system increases, since the gas cannot 
release energy during the process. Here it is important to note that 
the final temperature $T_2$ cannot be obtained from the well-known relation 
$P^{1-\gamma}T^{\gamma}=const.$, which is only valid if, in addition to being adiabatic, 
the process is quasi-static \cite{Cengel,vanWylen}, and clearly an abrupt compression 
does not satisfy the second condition. To derive $T_2$, we start from the first law for 
an adiabatic process:
	\begin{equation}
	\Delta U_{1\rightarrow 2}^{A}=-W^{A}.
	\end{equation}
Using that for a perfect gas $\Delta U_{1\rightarrow 2}^{A}=nC_V(T_2-T_1)$, and noting that 
the gas is compressed by a constant external pressure $P_2$ (although the gas pressure 
is probably not well-defined), we obtain that ($n=1$ mol):
	\begin{equation}\label{int1}
	C_V(T_2-T_1)=-P_2(V_2-V_1)
	\end{equation}
We can express the volumes in terms of the temperatures employing the ideal gas equation:
	\begin{equation}\label{int2}
		\begin{cases}
		V_{1}=RT_1/P_{1}\\
		V_{2}=RT_2/P_2,
		\end{cases}
		\end{equation}
so from Eqs. (\ref{int1}) and (\ref{int2}) we obtain:
	\begin{equation}
	C_V(T_2-T_1)=-RT_2+\dfrac{P_2RT_1}{P_1}.
	\end{equation}
Finally, defining $x$ as in Eq. (\ref{x}) and using that for an ideal gas $C_{P}-C_{V}=R$, 
after some work we obtain that the temperature of the equilibrium state is
	\begin{equation}\label{T2}
	T_{2}=T_{1}\left(1+\beta x\right),
	\end{equation}
where $\beta=R/C_{P}$.  Since the gas does not exchange heat with the environment, 
in this case the total entropy production is simply the entropy variation of the gas, 
which, as in the previous case, can be obtained integrating the Gibbs relation 
(\ref{Gibbs}) along a convenient trajectory. The general result is
	\begin{equation}\label{DSgeneral}
	\Delta S^{A}_{1\rightarrow 2}=C_P\ln\left(\dfrac{T_2}{T_1}\right)-R\ln\left(\dfrac{P_2}{P_1}\right),
	\end{equation}	 
so, finally, from Eqs. (\ref{x}), (\ref{T2}) and (\ref{DSgeneral}), we obtain:
	\begin{equation}\label{DS_univ_comp2}
	\Delta S^{\text{Univ}}_{1\rightarrow 2}=\Delta S^{A}_{1\rightarrow 2}=C_P[\ln(1+\beta x)-\beta\ln(1+x)].
	\end{equation}
The non-negativity of the above expression is a consequence of Bernoulli's inequality:
	\begin{equation}\label{Bernoulli}
	(1+x)^\beta\leq 1+\beta x,
	\end{equation}
which is valid for $0<\beta<1$ and $x\geq -1$ (notice that $\beta = R/C_P$ and $x$ 
given by Eq. (\ref{x}) satisfy these conditions).

\begin{figure}[!h]
	{\includegraphics[scale=0.38, clip]{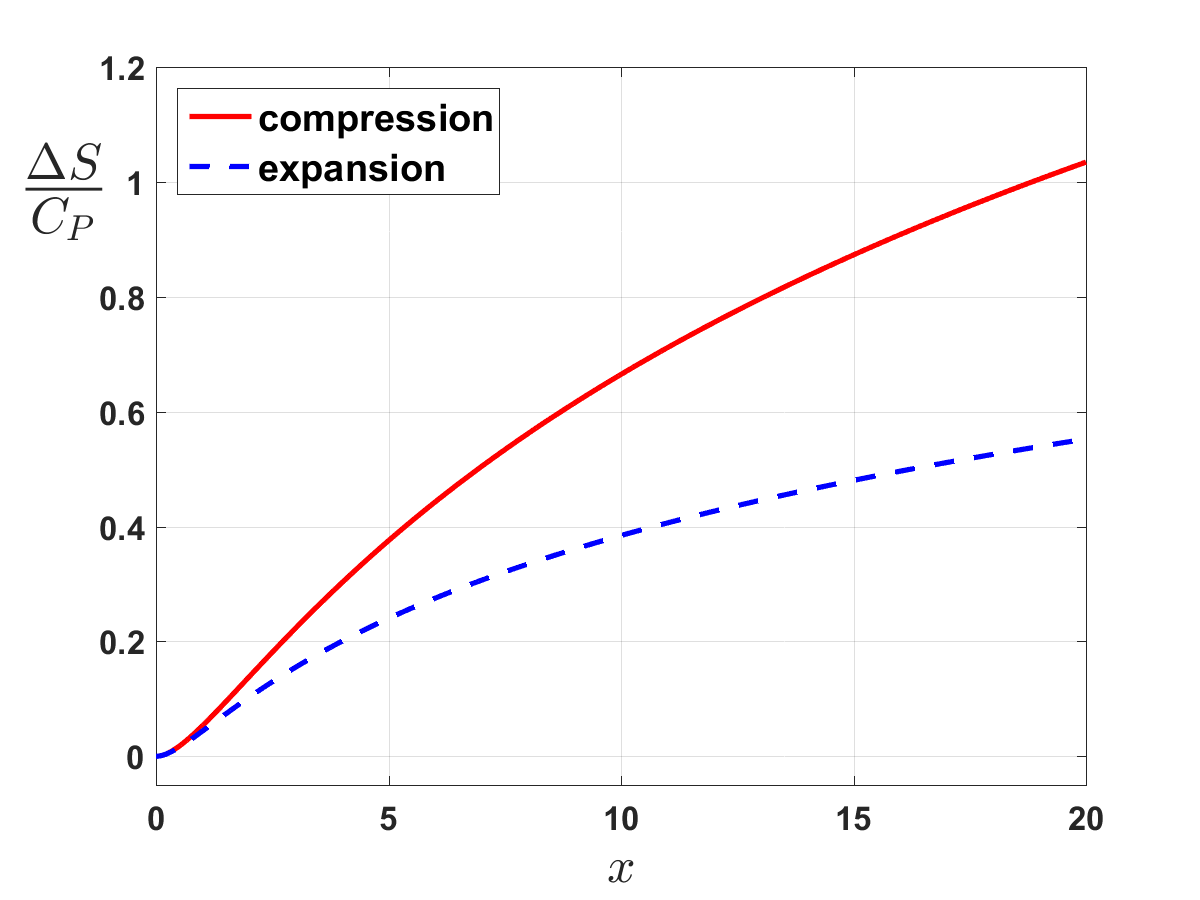}}
	\caption{Entropy production (in units of $C_P$) as a function of $x$, for the 
	processes of compression (red, solid line) and expansion (blue, dashed line) in 
	the adiabatic case. The working substance is nitrogen (diatomic gas, $\beta=2/7$) }
	\label{fig3}
\end{figure}

Now let us consider the expansion followed by the gas after suddenly changing the 
external pressure from the value $P_2$ to its initial value $P_{1}$. Following the 
same procedure, we find that the temperature of the final state $1'$ is
	\begin{equation}
	T_{1'}=T_{2}\left(\dfrac{1+x(1-\beta)}{1+x}\right),
	\end{equation}
so, from Eq. (\ref{DSgeneral}), the entropy created during the expansion adopts the 
form:
	\begin{equation}\label{DS_univ_exp2}
	\Delta S^{\text{Univ}}_{2\rightarrow 1'}=C_P\left[\beta\ln(1+x)-\ln\left(\dfrac{1+x}{1+x(1-\beta)}\right)\right].
	\end{equation}
In this case, another version of Bernoulli's inequality: 
	\begin{equation}
	\dfrac{1+x}{1+x(1-\beta)}\leq(1+x)^\beta,
	\end{equation}
allows us to derive the non-negativity of (\ref{DS_univ_exp2}). Notice that, unlike 
the case previously studied, the system does not return to the initial state when the 
pressure is suddenly reduced. This is due to the absence of heat exchange with the 
environment. However, the entropy productions (\ref{DS_univ_comp2}) and (\ref{DS_univ_exp2}) 
depend only on the pressures of the initial and final states, so they are independent on 
the order in which the processes are implemented. These quantities are represented in 
Fig. (\ref{fig3}) for the case of a diatomic gas.

Note that, unlike the previous case, both quantities present a logarithmic growth 
with the relative change of pressure in the limit of large $x$. For small values 
of $x$, Eqs. (\ref{DS_univ_comp2}) and (\ref{DS_univ_exp2}) adopt the approximate form

	\begin{equation}
	\Delta S^{\text{Univ}}_{1\rightarrow 2}\simeq\Delta S^{\text{Univ}}_{2\rightarrow 1'}= \dfrac{C_P\beta(1-\beta)}{2}\left(\dfrac{\Delta P}{P_1}\right)^2+\mathcal{O}\left(\dfrac{\Delta P}{P_1}\right)^3,
	\end{equation}
presenting a quadratic behaviour as in the previous case.

It is also interesting to notice that, for $\beta=1/2$, the expressions (\ref{DS_univ_comp2}) 
and (\ref{DS_univ_exp2}) reduce to
	\begin{equation}
	\Delta S^{\text{Univ}}_{1\rightarrow 2}=\Delta S^{\text{Univ}}_{2\rightarrow 1'}=C_p\left[\ln\left(1+\dfrac{x}{2}\right)-\dfrac{1}{2}\ln\left(1+x\right)\right],
	\end{equation}
so the compression and the expansion involve exactly the same level of irreversibility.  
This case corresponds to a heat capacity ratio
	\begin{equation}
	\gamma =\dfrac{C_p}{C_v}=\dfrac{1}{1-\beta}=2,
	\end{equation}
so recalling the relation between $\gamma$ and the degrees of freedom per particle $f$:
	\begin{equation}
	\gamma=1+\dfrac{2}{f},
	\end{equation}
we conclude that this symmetric behaviour of the entropy generations occurs only for $f=2$, 
i.e., for two-dimensional monoatomic gases. 

\begin{figure}[!h]
	{\includegraphics[scale=0.38, clip]{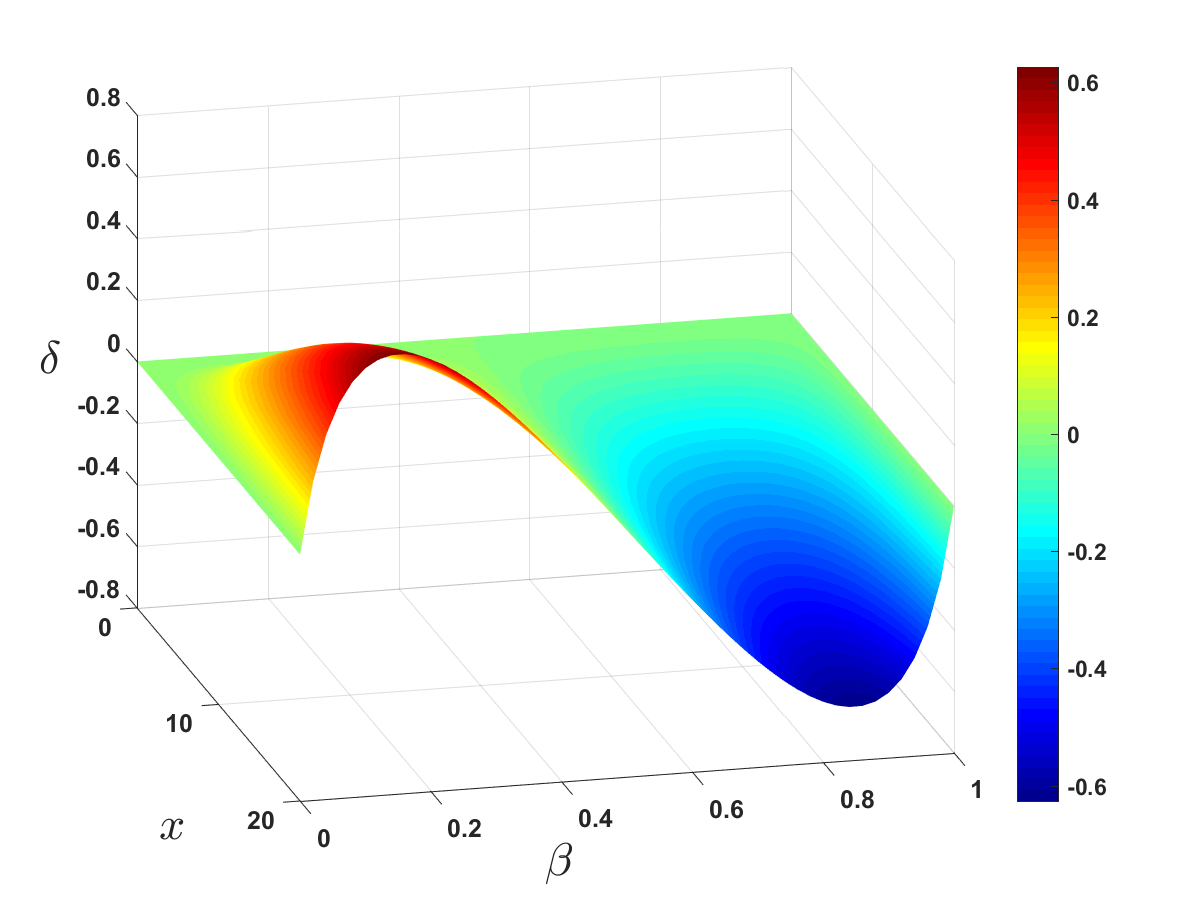}}
	\caption{Difference between the entropies produced during the sudden compression and 
	the sudden expansion, $\delta$, as a function of the relative change of pressure $x$ 
	and $\beta$.}
	\label{fig4}
\end{figure}

\begin{figure}[!h]
	{\includegraphics[scale=0.38, clip]{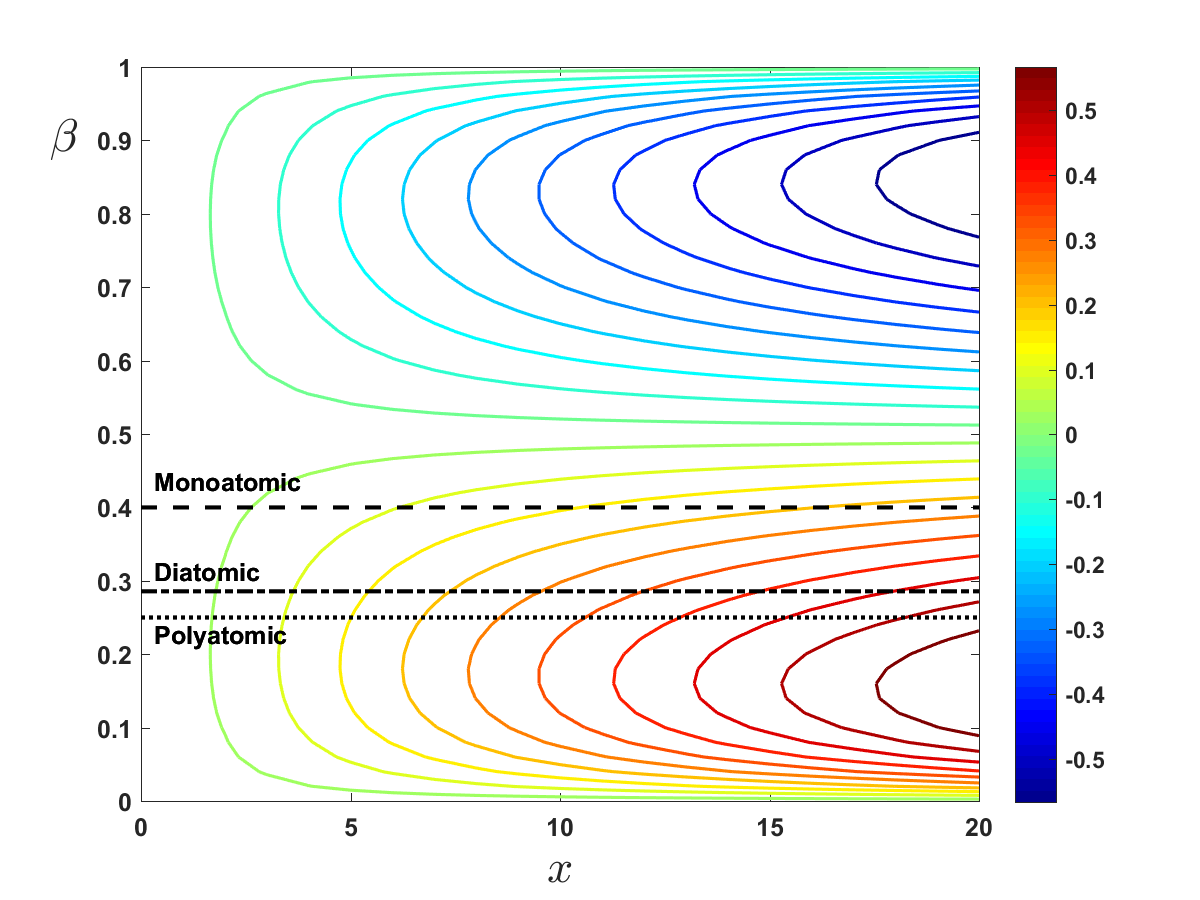}}
	\caption{Contour levels of the function $\delta(\beta,x)$ shown in Fig. (\ref{fig4}). 
	The dashed lines correspond to different types of perfect gases in three dimensions. 
	For those gases, the sudden compression always generates more entropy than the sudden 
	expansion.}
	\label{fig5}
\end{figure}

The curves in Fig. (\ref{fig3}) show the behaviour of entropy production for a diatomic gas, 
but similar results are obtained for other types of three-dimensional gases. To verify this statement, 
let us define the quantity  
	\begin{equation}
	\delta(\beta,x)=\dfrac{\Delta S^{\text{Univ}}_{1\rightarrow 2}-\Delta S^{\text{Univ}}_{2\rightarrow 1'}}{C_P},
	\end{equation}
whose sign allows us to infer which process is closer to the reversible limit. By virtue of Eqs. 
(\ref{DS_univ_comp2}) and (\ref{DS_univ_exp2}), $\delta$ adopts the form:
	\begin{equation}
	\delta(\beta,x)=\ln\left[\dfrac{(1+\beta x)(1+x)}{1+x(1-\beta)}\right]-2\beta\ln(1+x).
	\end{equation}
Since $\delta(\beta,x=0)=0$, the sign of $\delta$ is defined by the sign of its derivative with 
respect to $x$:
	\begin{equation}
	\dfrac{\partial\delta}{\partial x}=\dfrac{\beta(1-2\beta)(1-\beta)x^{2}}{(1+x)[1+x(1-\beta)](1+\beta x)},
	\end{equation}
which is positive for $0<\beta<1/2$. Since the maximum value of $\beta$ for a gas in three dimensions is 2/5 (monoatomic gas), we conclude that, 
despite the logarithmic growth shown in both processes, the compression is always further from the reversible limit than the expansion. 
The mentioned behaviour can be observed on Figs. (\ref{fig4}) and (\ref{fig5}).

\section{Remarks and conclusions}

Entropy production is a measure of the additional work that could have been 
obtained if the processes were carried out in a reversible manner. For that 
reason, the determination of the entropy production along given process, and
the comparison of processes from the entropy generation point of view
are matters of great practical interest. In this paper we have 
addressed this topic by performing the comparison between the processes 
of sudden compression and sudden expansion of a perfect gas.
The analysis was carried out both in the case in which the gas exchanges heat 
with a thermal reservoir and when it is thermally isolated. 

The results show 
that, in both cases, the compression involves a higher level of entropy generation
for all possible values of the initial and final pressures. In our experience 
this result is counter-intuitive for students, perhaps due to the use of the free 
expansion as a paradigmatic example of irreversible process in most textbooks. 

For the case in which the gas exchanges heat with a reservoir the comparison between both processes  
was performed geometrically, interpreting the entropy productions as 
areas in a certain diagram. This diagram is an interesting didactic tool, 
since it makes evident the strong asymmetry of the entropies produced in 
both processes, in the limit of large relative change of pressure. 
The mentioned asymmetry was also found (although in a lesser degree) when 
the system is isolated from the environment. 

Finally, it is important to highlight that the analysis has been performed for 
perfect gases, so the validity of the results is restricted to such kind of systems. 
For example, in the diathermal case, the fact that the internal energy of a perfect 
gas depends only on the temperature implies that we can found its entropy variation 
integrating the function $f(P)=R/P$, and the positive concavity of this function 
determines that the compression creates more entropy (see Fig. (\ref{fig1})). However, for high pressures 
(or low temperatures), the interaction 
between the molecules of the gas is not negligible, and the ideal gas model, on which 
the deductions are based, starts to fail. As an example, let us consider \SI{1}{mol} 
of water steam in
an adiabatic cylinder-piston device at \SI{6}{MPa} and \SI{450}{\celsius} 
(compressibility factor $Z\simeq$ 0.93) that is is abruptly 
compressed to \SI{7}{MPa} by changing the weight over the piston. Using tables of thermodynamic 
properties usually employed in undergraduate courses \cite{Cengel}, it is possible to show 
that the entropy increase of the water 
is  $\Delta S_{1\rightarrow 2}\simeq$\SI{0.0028}{J/K}. On the other hand, after the removal of
the weight, the entropy increase associated to the expansion process (back to the initial 
pressure of \SI{6}{MPa}) is 
$\Delta S_{2\rightarrow 1'}\simeq$\SI{0.0049}{J/K}, so it exceeds the previous value. 
This analysis shows that the conclusions 
obtained in this work cannot be extrapolated to real gases far from ideal behaviour.

We believe that, through the trigger question about which process is further from 
the reversible limit, the results presented here can be used to plan a 
didactic activity that could improve the understanding of the 
irreversibility concept, exercise entropy analysis as well as 
develop thermodynamic intuition in students.

\section*{Acknowledgments}
This work was partially supported by Agencia Nacional de Investigación e Innovación (Uruguay). 
The authors thank the anonymous referees for their comments that helped to improve the manuscript.

\end{document}